\documentclass[aps,pra,twocolumn,groupedaddress,showpacs,10pt]{revtex4-1}

\usepackage{amsmath,amsfonts,amssymb,bbm,graphicx}

\newcommand{\eqn}[1]{\begin{equation} #1 \end{equation}} 
\newcommand{\aln}[1]{\begin{align} #1 \end{align}}       
\newcommand{\mul}[1]{\begin{multline} #1 \end{multline}} 
\newcommand{\mc}{\mathcal}                               
\newcommand{\mbf}{\mathbf}                               
\newcommand{\bs}{\boldsymbol}                
\newcommand{\eq}[1]{(\ref{#1})}              
\newcommand{\pd}{\partial}                   
\renewcommand{\l}{\left}
\renewcommand{\r}{\right}
\newcommand{\Tr}{\text{Tr}\,}                
\newcommand{\q}{\text q}                     
\newcommand{\cl}{\text {cl}}
\newcommand{\A}{\text A}
\newcommand{\R}{\text R}
\newcommand{\K}{\text K}

\begin{document}


\title{Transmission statistics in nonconservative disordered optical medium}


\author{Zhong Yuan Lai}
\affiliation{Physikalisches Institut, Universit\"at Bonn, Nu{\ss}allee 12, 53115 Bonn, Germany}

\author{Oleg Zaitsev}
\affiliation{Physikalisches Institut, Universit\"at Bonn, Nu{\ss}allee 12, 53115 Bonn, Germany}



\begin{abstract}

We determine the cumulants of electromagnetic energy transmitted through one-dimensional disordered medium with absorption or amplification. For this purpose we derive the Keldysh nonlinear $\sigma$ model action with a source term that generates the energy current fluctuations. The fluctuations over the ensemble of disorder realizations are found to decrease (increase) with increasing absorption (amplification). In the conservative medium the fluctuations can be related to the Dorokhov's distribution of transmission coefficients.

\end{abstract}

\pacs{42.25.Dd}

\maketitle


\section{Introduction}

Physics of disordered mesoscopic conductors~\cite{blan00,been03} made a strong impact on the research on light propagation in disordered media because of the well-known analogy between the scalar wave equation and the Schr\"{o}dinger equation. Important information about the system can be gained by studying fluctuations of charge current (or energy current in optical medium), which are usually described via the \emph{current cumulants}~\cite{naza09}. The calculation of these quantities normally requires the knowledge of the full statistical distribution of transmission eigenvalues~\cite{naza03, lern04}.

The theory for electronic charge transmission has a long history and is well developed. Various methods were employed to compute the full counting statistics of electronic conduction through disordered wires, among them the semiclassical formalism~\cite{jong96}, random matrix theory~\cite{been92}, and the Green's function formalism~\cite{naza94}. In particular, it was found in all these works that the electronic shot noise in disordered conductors is reduced below the Poissonian (uncorrelated) value by a universal factor of $\frac{1}{3}$. This reduction is a result of the existence in a disordered conductor of a fraction of non-weakly transmitting (``open'') channels with transmission coefficients close to~$1$, with the remainder being weakly conducting (``closed'') channels with small transmission coefficients. This property is reflected in the distribution of transmission coefficients originally derived by Dorokhov~\cite{doro84}. Although shot noise is not present in the transmission of classical electromagnetic waves, the transmission statistics is still governed by the Dorokhov's distribution, as we show below. 

In the field of optics, transmission statistics in disordered channels was extensively studied experimentally~\cite{gena87,vell08,popo10,shi12,davy13} and slightly less theoretically~\cite{debo94,vanl96}. We are not aware of similar research done on dissipative or amplifying systems, which can be relevant for practical applications, such as random lasers~\cite{cao05}.

In the present work we rederive the Keldysh nonlinear $\sigma$ model for nonconservative medium~\cite{lai12} in the form that allows one to calculate the energy current fluctuations. Specifically, we include the source term in the action and follow the steps outlined in Ref.~\cite{kame11}. In the stationary-phase approximation one obtains the Usadel equation which admits an exact solution for a one-dimensional conservative system~\cite{kame11,naza94}. We solve the Usadel equation in the weakly nonconservative system by a perturbation expansion and calculate the action at the stationary point. The action generates the cumulants of the energy transmitted during a fixed time interval. We apply the general theory in the special cases of thermal fluctuations in equilibrium and fluctuations over disorder realizations in the pumped system at zero temperature. The latter are related to the fluctuations of the transmission coefficients described by the Dorokhov's distribution. We also find quantitative agreement with previous diagrammatic calculations~\cite{debo94} of the second- and third-order cumulants in the conservative medium.

\section{Nonlinear sigma model with source}

\subsection{Keldysh field theory}

Let us consider a classical electromagnetic wave with transverse magnetic (TM) polarization in a two-dimensional medium. Following Ref.~\cite{lai12}, we introduce two complex fields, $A_\omega (\mbf r)$ and~$A_\omega^* (\mbf r)$, whose real parts correspond to the normal component of vector potential at the position~$\mbf r$ and frequency~$\omega$. The physically relevant frequencies for the fields are restricted to the neighborhood of the typical optical frequency $\omega = \omega_0 > 0$, which is assumed to be the largest frequency scale in the system.

A nonconservative medium can be conveniently described within the Keldysh formalism~\cite{keld64a,*keld64b}. Here one considers the system's evolution first forward and then backward in time. Correspondingly, each field acquires two components denoted as $A_\pm$ and $A_\pm^\dag$ for positive ($+$) and negative ($-$) time direction. [Here and below functions without (some of the) arguments are considered as column or row vectors in the appropriate Hilbert space.] The Keldysh rotation
\eqn{
  A^\cl = \frac 1 {\sqrt 2} (A_+ + A_-), \quad
  A^\q = \frac 1 {\sqrt 2} (A_+ - A_-)
\label{rot}
}
defines the classical and quantum field components which form a vector
\eqn{
  \hat A = \l(\begin{array}{cc} A^\cl \\ A^\q \end{array} \r)
}
in the Keldysh space of twice the dimensionality of the original Hilbert space.

The basic object of the Keldysh field theory~\cite{kame11} is the functional-integral form of the partition function,
\eqn{
  Z = \int\! D[\hat A, \hat A^\dag]\, e^{iS [\hat A, \hat A^\dag]},
  \quad \hbar = 1,
\label{Z}
}
with the measure
\eqn{
  D[\hat A, \hat A^\dag] = \mc N  \prod_{\substack{\mbf r, \omega,\\
  j = \cl, \q}} \frac {d \bigl(\text{Re}\, A^j_\omega (\mbf r) \bigr)\,
  d \bigl(\text{Im}\, A^j_\omega (\mbf r) \bigr)} \pi.
\label{DA}
}
All information about the system is contained in the action
\eqn{
  S [\hat A, \hat A^\dag] = \frac 1 {16 \pi} \hat A^\dag\, \hat G^{-1} \hat A,
\label{S0}
}
where the inverse Green's function operator has the form
\eqn{
  \hat G^{-1} = \l(\begin{array}{cc} 0 & (G^{-1})^\A \\ (G^{-1})^\R &
  (G^{-1})^\K \end{array} \r)
\label{G-1_matr}
}
(we remind that matrix products involve integration over continuous variables). The retarded, advanced, and Keldysh components of $\hat G^{-1}$ are given by (we set the velocity of light $c = 1$)
\aln{
  &(G^{-1})^{\R,\A}_\omega (\mbf r) = \epsilon' (\mbf r, \omega)\, \omega^2
  + \mbf \pd_{\mbf r}^2 \pm i \epsilon'' (\mbf r, \omega)\, \omega^2,
\label{G-1_RA}\\
  &(G^{-1})^\K = (G^{-1})^\R F_0 - F_0\, (G^{-1})^\A,
\label{G-1_K}\\
  &\epsilon' (\mbf r, \omega) \equiv \text{Re}\,
  [\epsilon (\mbf r, \omega)], \quad
  \epsilon'' (\mbf r, \omega) \equiv \text{Im}\, [\epsilon (\mbf r, \omega)],
}
where $\epsilon (\mbf r, \omega)$ is the dielectric constant of the medium and
\eqn{
  F_0 = 2 n_0 + 1 = \coth \frac \omega {2 T} \quad (k_{\text B} = 1)
\label{F0}
}
is related to the photon occupation number~$n_0$, which is given here for the thermal equilibrium at temperature~$T$. The normalization constant~$\mc N$ in Eq.~\eq{DA} depends on the discretization of continuous variables and ensures the property $Z = 1$, which reflects the fact that the system arrives to its initial state as a result of the forward-backward evolution.

In order to compute averages of physical observables, suitable ``source'' terms are added to the action~\eq{S0}. The source term for the energy current density is (cf.\ Ref.~\cite{kame11})
\eqn{
  \Delta S = \Tr \l( \bs \lambda \cdot \mbf j^\cl \r),
\label{DeltaS}
}
where $\mbf j^\cl$ is the classical field component of the current
\eqn{
  \mbf j (\mbf r, t) = -\frac 1 {16 \pi} \l[ \l( \pd_{\mbf r}
  A^* \r) \l( \pd_t A \r) + \l( \pd_t A^* \r) \l( \pd_{\mbf r} A \r) \r]
}
and the source field $\bs \lambda (\mbf r, t)$ is the quantum component of the field $(0, \bs \lambda)^{\text T}$ with the zero classical component in the Keldysh space. Here and below the trace includes integration over continuous variables. The average current is then given by the functional derivative of the (logarithm of) partition function with respect to the source,
\eqn{
  \langle \mbf j (\mbf r, t) \rangle = -i \frac {\delta \ln Z [\bs \lambda]}
  {\delta \bs \lambda (\mbf r, t)}\biggr|_{\bs \lambda \equiv 0}
  = -i \frac {\delta Z [\bs \lambda]} {\delta \bs \lambda
  (\mbf r, t)}\biggr|_{\bs \lambda \equiv 0}.
}
The fact that $\bs \lambda$ is a quantum field is responsible for $Z [\bs \lambda] \ne 1$, in general, whereas $Z [\bs \lambda = 0] = 1$ by construction.

Performing the Fourier transform we approximate $\pd_t A \mapsto -i \omega A \approx -i \omega_0 A$ and $\pd_{\mbf r} A \mapsto i \mbf k A \approx i \omega_0 \sqrt{\epsilon'} \bs \kappa (\mbf k) A$, where $\bs \kappa (\mbf k) = \mbf k/|\mbf k|$. Then the sum of contributions~\eq{S0} and~\eq{DeltaS} yields the total action
\eqn{
  S [\hat A, \hat A^\dag, \bs \lambda] = \frac 1 {16 \pi} \hat A^\dag \l(
  \hat G^{-1} + \sqrt{2 \epsilon'} \omega_0^2 \bs \kappa \cdot \bs \lambda \hat   1 \r) \hat A,
\label{S_lambda}
}
where $\hat 1$ is the unit operator in the Keldysh space.

\subsection{Nonlinear sigma model}

We neglect the dispersion of the dielectric constant
\eqn{
  \epsilon (\mbf r, \omega) \simeq \epsilon' + \Delta \epsilon' (\mbf r) +
  i \epsilon''
}
and assume that its real part has a random component with $\langle \Delta \epsilon' (\mbf r) \rangle = 0$ and
\eqn{
  \langle \Delta \epsilon' (\mbf r)\, \Delta \epsilon' (\mbf r') \rangle =
  \frac {2 {\epsilon'}^2} {\pi \nu_0 \omega_0^2 \tau} \,
  \delta (\mbf r - \mbf r'),
}
where $\nu_0$ is the optical density of modes at the frequency~$\omega_0$, $\tau$~is the scattering time, and the averages are taken over disorder realizations. Disorder-averaged partition function [which generates the cumulant expansion \eq{cumexp} below and eventually enables us to calculate the cumulants]
\aln{
  &\langle Z [\bs \lambda] \rangle = \int D [\hat Q] \, e^{iS [\hat Q,
  \bs \lambda]},
\label{Zdis}\\
  &i S[\hat Q, \bs \lambda] = - \Tr \l[ \frac {\pi \nu_0}
  {4 \tau} \hat Q^2 \vphantom{\frac {\epsilon' \omega_0} \tau}\r. \notag\\
  &\phantom{i S[\hat Q, \bs \lambda]} + \l.\ln \l(\hat G_0^{-1} +
  \frac {\epsilon' \omega_0} \tau \hat \gamma \hat Q + \sqrt{2 \epsilon'}
  \omega_0^2 \bs \kappa \cdot \bs \lambda \hat 1 \r) \r],
\label{SQ}\\
  &\hat \gamma \equiv \l( \begin{array}{cc} 0 & 1 \\ 1 & 0 \end{array} \r),
}
can be expressed via the functional integral over the field~$\hat Q$ which has a $2 \times 2$ matrix structure in the Keldysh space; here $\hat G_0^{-1}$ is the inverse Green's function operator in a medium with dielectric constant~$\epsilon'$ without disorder. The random~$\Delta \epsilon' (\mbf r)$ was neglected in the source term of Eq.~\eq{S_lambda}, because its contribution to $S[\hat Q, \bs \lambda]$ would be small compared to the disorder-free source term in the limit $\omega_0 \tau \gg 1$.

The stationary point of the action~\eq{SQ} has the form
\eqn{
  \hat {\underline Q} = i \hat \Lambda, \quad
  \hat \Lambda = \l(\begin{array}{l@{\hspace{10pt}}c} 1^\R & 1^\R F +
  F\, 1^\A \\ 0 & -1^\A \end{array} \r).
\label{Lambda}
}
Here the operator~$F$, the so called distribution function, is to be determined later. In thermal equilibrium, $F=F_0$ [Eq.~\eq{F0}] is uniform in space. The retarded and advanced unit operators, $1^{\R,\A}_\omega = e^{\pm i \omega \varepsilon}$, where $\varepsilon \ll \omega_0^{-1}$ is an arbitrary constant, oscillate at high frequencies $\omega \gtrsim \varepsilon^{-1}$, which leads to the property \mbox{$\Tr \hat {\underline Q}^2 = 0$}. Note that $\hat {\underline Q}^2 = - \hat 1$.

The dominant contribution to~$\langle Z [\bs \lambda] \rangle$ arises from
the trace-preserving fluctuations of $\hat Q$ about the stationary point that satisfy the conditions
\eqn{
  \Tr \hat Q^2 = 0, \quad \hat Q^2 = - \hat 1.
\label{TrQ}
}
Fluctuations of this type, the ``massless'' modes, do not affect the $\hat Q^2$~term in~$S[\hat Q, \bs \lambda]$, and result in a weaker variation of the action compared to arbitrary, ``massive'' fluctuations. The massless modes describe the diffusive light propagation. Following the general prescription in Ref.~\cite{kame11}, we generalize the expression for the effective action of massless modes~\cite{lai12} in the presence of source:
\aln{
  &i S[\hat Q, \bs \lambda] \simeq \notag \\
  &- \pi \nu_0 \Tr\! \l[-i \pd_t \hat Q + \frac D 4
  (\hat \pd_{\mbf r} \hat Q)^2 - \frac i {2 \tau_{\text a}}
  \hat Q \hat \Lambda_0 \r],
\label{Seff}
}
where $(\pd_t \hat Q)_{tt} \equiv (\pd_t \hat Q_{tt'} )_{t'=t}$,
\eqn{
  D = \frac \tau {2 \epsilon'}
}
is the diffusion coefficient, the covariant derivative is defined~as
\eqn{
  \hat \pd_{\mbf r} \hat Q = \pd_{\mbf r} \hat Q - i \frac {\omega_0} {\sqrt 2}
  [\bs \lambda \hat \gamma, \hat Q]_-
}
($[\, , ]_-$ being the commutator), the absorption rate (negative for amplification)~is
\eqn{
  \frac 1 {\tau_{\text a}} = \frac {\epsilon'' \omega_0} {\epsilon'},
}
and $\hat \Lambda_0$ is given by $\hat \Lambda$ [Eq.~\eq{Lambda}] with $F = F_0$. We note that the disorder-free inverse Green function can be written~as
\eqn{
  \hat \gamma \hat G_0^{-1} = (\epsilon' \omega^2 + \pd_{\mbf r}^2)\, \hat 1
  + i \epsilon'' \omega_0^2 \, \hat \Lambda_0,
}
i.e., $\hat \Lambda_0$ describes the thermal bath responsible for dissipation (if $\epsilon'' > 0$)~\footnote{In Ref.~\cite{lai12} we did not discriminate between $\hat \Lambda$ and~$\hat \Lambda_0$, hence, the results of this reference are valid only in the case of thermal equilibrium.}. The $\hat Q$-independent contributions to the action are omitted. The approximation~\eq{Seff}, which bears the name of \emph{nonlinear $\sigma$~model}, is derived under the assumptions of weak variation of~$\hat Q$ in time and space and weak absorption,
\eqn{
  \frac \tau {|\tau_{\text a}|} \ll 1.
}
Corrections to the diffusion coefficient~$D$ due to absorption are neglected within this order of approximation (see the discussion in Ref.~\cite{lai12}).

\section{Transmission statistics}

\subsection{Stationary-phase approximation}

We apply the general method to compute fluctuatons of energy transmitted through a quasi-one-dimensional disordered system. Let the light be continuously pumped into the channel at one end, $x = 0$, and the outgoing energy $E = t_0\, j(L)$ accumulated during a time interval~$t_0$ be measured at the other end, $x = L$. [In the stationary regime the one-dimensional current~$j(x)$ has no time dependence.] The source field
\aln{
  &\lambda (x,t) = \notag \\
  &\l\{
    \begin{array}{ll}
       \frac \eta {\sqrt 2 l}, & L - l \leq x \leq L \text{ and }
       0 \leq t \leq t_0 \\
       0, & \text{otherwise}
     \end{array}
   \r., \quad l \to 0,
}
is constructed in such a way that it couples to the current averaged over a narrow interval of length~$l$ near the channel's end during the time~$t_0$~\footnote{Note that $j^\cl / \sqrt 2 = (j_+ + j_-)/2$}. This form of the source field generalizes the uniform field of Ref.~\cite{kame11}, which was sufficient for a lossless medium where $j(x) = \text{const}$. The disorder-averaged partition function $\langle Z(\eta) \rangle$ defines the cumulants $C_m$ of energy~$E$ via the expansion of its logarithm in the counting variable~$\eta$:
\eqn{
  \ln\, \langle Z(\eta) \rangle = \sum_{m=1}^\infty \frac {(i \eta)^m}
  {m!}\, C_m.
\label{cumexp}
}
In particular, $C_1 = \langle E \rangle$ and for $m = 2, 3$ the cumulants are equal to the central moments, $C_m = \bigl< (E - \langle E \rangle)^m \bigr>$. For $m \geq 4$ the $m$th central moment can be expressed in terms of $C_{m'}$'s with $m' \leq m$.

With the help of the gauge transformation
\aln{
  &\hat Q_\eta (x) = \l\{
    \begin{array}{ll}
      \hat Q (x), & 0 \leq x \leq L - l \\
      e^{-i \alpha (x) \hat \gamma}\, \hat Q (x)\,
      e^{i \alpha (x) \hat \gamma},
      & L - l \leq x \leq L
    \end{array}
  \r., \\
  &\alpha (x) =  \frac {x - L + l} {2l} \, \omega_0 \eta, \quad
  \alpha \equiv \alpha (L) = \frac {\omega_0} 2 \eta,
}
having the property $\hat \pd_x \hat Q = \pd_x \hat Q_\eta$, the explicit source contribution is eliminated from the action~\eq{Seff}, which now reads
\mul{
  i S [\hat Q_\eta] = \\
  - \pi \nu_0 \Tr\! \l[-i \pd_t \hat Q_\eta + \frac D 4  (\pd_{\mbf r} \hat
  Q_\eta)^2  - \frac i {2 \tau_{\text a}}  \hat Q_\eta (\hat \Lambda_0)_\eta
  \r].
}
By allowing $\hat Q_\eta$ to fluctuate under the constraints~\eq{TrQ} and setting the linear variation of~$S [\hat Q_\eta]$ to zero we obtain the Usadel equation
\mul{
  -i (\pd_t + \pd_{t'}) \bigl(\hat {\underline Q}_\eta \bigr)_{tt'} +
  D\, \pd_x \bigl(\hat {\underline Q}_\eta \pd_x
  \hat {\underline Q}_\eta \bigr) \\
  - \frac i {2 \tau_{\text a}} \bigl[(\hat \Lambda_0)_\eta, \, \hat {\underline
  Q}_\eta \bigr]_-  = 0
}
for the stationary-point configuration~$\hat {\underline Q}_\eta$. Without the source ($\eta = 0$), the Usadel equation reduces to the equation
\eqn{
  \bigl(- \pd_t + D \pd_x^2 \bigr) F - \tau_{\text a}^{-1} (F - F_0) = 0
\label{difeq}
}
for the distribution function~$F_\omega (x,t)$ obtained from the matrix $F_{t't''} (x)$ by making the Fourier transform in the fast variable $t' - t''$ and keeping the slow variable $t = (t' + t'')/2$.

In general, $\hat {\underline Q}_\eta$~is not of the form~\eq{Lambda}. The source enters the Usadel equation via the boundary conditions. We will model the pump at $x = 0$ by the time-independent non-equilibrium distribution function within the frequency band of width $\Delta \omega \ll \omega_0$:
\eqn{
  F_\omega (0) = \l\{
    \begin{array}{ll}
      F_*, & |\omega - \omega_0| < \Delta \omega/2 \\
      F_0, & |\omega - \omega_0| > \Delta \omega/2
    \end{array}
  \r.,
}
where $F_* \geq F_0$ and $F_0$ is the equilibrium distribution~\eq{F0} at $\omega = \omega_0$. At the other end, $x = L$, the channel is assumed to be in contact with the thermal bath. Thus, the boundary conditions are
\aln{
  &\hat {\underline Q}_\eta (0) = \hat {\underline Q} (0) = i \Lambda_*, \\
  &\hat {\underline Q}_\eta (L) = i (\Lambda_0)_\eta (L) =
  i e^{-i \alpha \hat \gamma} \Lambda_0\, e^{i \alpha \hat \gamma},
\label{bc}
}
where $\Lambda_*$ depends on~$F_\omega (0)$.

We evaluate the functional integral for~$\langle Z (\eta) \rangle$ by the stationary-phase approximation:
\eqn{
  \langle Z (\eta) \rangle \simeq a (\eta)\, e^{i S (\eta)}
  \simeq  e^{i S (\eta)}, \quad
  S (\eta) \equiv S \bigl[\hat {\underline Q}_\eta \bigr],
}
where we neglect the $\eta$ dependence of the prefactor and the normalization $\langle Z (0) \rangle = 1$ is guaranteed by the form of~$\hat {\underline Q}_{\eta = 0}$~\eq{Lambda}. Thus, to obtain the cumulants we need to find the time-independent solution of the Usadel equation with boundary conditions~\eq{bc}, and calculate the action~$S (\eta)$.

\subsection{Weakly nonconservative medium}

We solve the Usadel equation in the limit of small absorption or amplification retaining only the leading order in the small parameter~$L^2/l_{\text a}^2$, where the absorption length squared (negative for amplification) is defined~as
\eqn{
  l_{\text a}^2 = D \, \tau_{\text a}.
}
The second-order Usadel equation is equivalent to the system of two first-order equations:
\aln{
  &\pd_x \hat J = \frac i {2 \tau_{\text a}} \bigl[(\hat \Lambda_0)_\eta, \,
  \hat {\underline Q}_\eta \bigr]_-,
\label{J}  \\
  &\hat J = \hat {\underline Q}_\eta \pd_x \hat {\underline Q}_\eta \;
  \l[ = - (\pd_x \hat {\underline Q}_\eta) \, \hat {\underline Q}_\eta \r],
\label{Q}
}
where the last equality follows from the property $\pd_x \bigl(\hat {\underline Q}_\eta^2 \bigr) = 0$. Equation~\eq{Q} can be formally integrated to yield
\eqn{
  \hat {\underline Q}_\eta (x) = \hat {\underline Q}_\eta (0)
  \l[ e^{\int_0^x dx' \hat J (x')} \r],
\label{Qint}
}
where the exponential is ordered in $x'$ increasingly from left to right. In the zeroth order in absorption one finds $\hat J (x) = \text{const}$, which leads to~\cite{kame11}
\aln{
  &\hat {\underline Q}_\eta^{(0)} (x) = \hat {\underline Q}_\eta (0)\,
  e^{\hat J_0 x}, \\
  &\hat J_0 \equiv L^{-1} \ln \bigl[- \hat {\underline Q}_\eta (0) \,
  \hat {\underline Q}_\eta (L) \bigr].
\label{J0}
}
In the next order we write $\hat J (x) = \hat J_0 + \Delta \hat J (x)$ and $\hat {\underline Q}_\eta (x) = \hat {\underline Q}_\eta^{(0)} (x) + \hat {\underline Q}_\eta^{(1)} (x)$. By expanding the ordered exponential in Eq.~\eq{Qint} in $\Delta \hat J (x)$ we obtain
\aln{
  \hat {\underline Q}_\eta^{(1)} (x)
  \simeq \hat {\underline Q}_\eta (0) \int _0^x d x'\,
  e^{\hat J_0 x'} \Delta \hat J (x')\, e^{\hat J_0 (x - x')}.
}
The requirement $\hat {\underline Q}_\eta^{(1)} (L) = 0$, that follows from the boundary conditions, yields the property
\eqn{
  \int _0^L d x\, e^{\hat J_0 x}\, \Delta \hat J (x)\, e^{-\hat J_0 x} = 0.
\label{DeltaJ}
}
The gradient contribution to the action contains
\aln{
  \Tr \bigl(\pd_x \hat {\underline Q}_\eta \bigr)^2 &= - \Tr \l[\bigl(\pd_x
  \hat {\underline Q}_\eta  \bigr)\, \hat {\underline Q}_\eta^2 \,
  \pd_x \hat {\underline Q}_\eta \r] = \Tr \hat J^2 \notag \\
  &\simeq \Tr \hat J_0^2 + 2 \Tr \bigl( \hat J_0 \Delta \hat J \bigr)
  = \Tr \hat J_0^2,
}
according to the property~\eq{DeltaJ}. Thus, $\Delta \hat J (x)$ does not contribute to the action, which now takes the form
\eqn{
  i S(\eta) \simeq - \pi \nu_0 \Tr\! \l[\frac D 4 \hat J_0^2 - \frac i
  {2 \tau_{\text a}} (\hat \Lambda_0)_\eta \,
  \hat {\underline Q}_\eta^{(0)} \r].
\label{S_eta}
}
The nonconcervative part of the action can be simplified in the limit $l \to 0$ by tracing over~$x$ explicitly and disregarding the interval $L - l \leq x \leq L$; the remaining integral can be evaluated with $l = 0$ (note that $\hat J_0$ does not depend on~$l$) yielding
\eqn{
  i S_{\text{nc}} (\eta) = \frac {i \pi \nu_0} {2 \tau_{\text a}} \,
  \Tr_{\omega,\K}\! \l[ \hat \Lambda_0\, \hat {\underline Q} (0)\,
  \hat J_0^{-1} \l(e^{\hat J_0 L} - \hat 1 \r) \r],
\label{Snc}
}
where the trace is performed in the $\omega$ and Keldysh subspaces. Below we apply the general expressions in the cases of thermal equilibrium and transport at zero temperature.

\begin{widetext}

\subsection{Special cases}

\subsubsection{Thermal fluctuations}

We consider the fluctuations of transmitted energy in the absence of pumping, $F_* = F_0$, at finite temperature~$T$. The leading contribution to the action~\eq{S_eta}~is
\eqn{
  i S_0 (\eta) = - \frac {\pi \nu_0 D t_0} {4L} \int_0^\infty \!
  \frac {d \omega} {2 \pi} \, 2 (\ln \lambda)^2,
\label{TrJ0}
}
where $\lambda$ is one of the two eigenvalues
\aln{
  &\lambda_{1,2} = 1 + X \pm \sqrt{X (X + 2)}, \quad \lambda_1 \lambda_2 = 1,\\
  &X \equiv 2 \, (F_0^2 - 1) \, \sin^2 \alpha,
}
of the matrix $\hat \Lambda_0 \, e^{-i \alpha \hat \gamma} \hat \Lambda_0 \, e^{i \alpha \hat \gamma}$ appearing in the logarithm in Eq.~\eq{J0}. According to Eq.~\eq{F0} only frequencies $\omega \lesssim T$ contribute to the integral~\eq{TrJ0}~\footnote{To regularize the integral at low frequencies we should avoid fixing~$\omega$ in $\alpha = \omega \eta/2$.}. The nonconcervative contribution reads
\aln{
  i S_{\text{nc}} (\eta) &= - \frac {\pi \nu_0} {2 \tau_{\text a}} t_0 L
  \int_0^\infty \! \frac {d \omega} {2 \pi} \, 2 \l\{ \frac {\sqrt{X (X + 2)}}
  {\ln \lambda_1} - 1 \r\} \notag \\
  &= \frac {\pi \nu_0} {2 \tau_{\text a}} t_0 L \int_0^\infty \! 
  \frac {d \omega} \pi \l\{ \frac {\sqrt{\l[ 1 + 2 \, (F_0^2 - 1) \, 
  \sin^2 \alpha \r]^2 - 1 }} 
  {\ln \l[ 1 + 2 \, (F_0^2 - 1) \, \sin^2 \alpha + \sqrt{\l[ 1 + 2 \, 
  (F_0^2 - 1) \, \sin^2 \alpha \r]^2 - 1} \r]} - 1 \r\}, \quad
  \alpha = \frac {\omega_0} 2 \eta,
\label{SncT}
}
\end{widetext}
where the unity in the braces is subtracted in order to account for the high-frequency regularization due to~$1^{\R,\A}$. To justify this result, let us, first, set $T = 0$ ($F_0 = 1$). In this case the matrix~$\hat J_0$ cannot be diagonalized, but can be brought to the upper triangular form with $1^{\R,\A}$ on the diagonal by the rotation $\hat {\mc P}^{-1} \hat J_0 \hat {\mc P}$, where
\eqn{
  \hat {\mc P} = \frac 1 {\sqrt 2} \l( \begin{array}{rr} 1 & 1 \\ -1 & 1
  \end{array} \r), \quad \hat {\mc P}^{-1} = \hat {\mc P}^{\text T}.
}
The regularization makes the trace in Eq.~\eq{Snc} vanish, which results in $S_{\text{nc}} (\eta; T = 0) = 0$. For finite temperature we can write
\aln{
  S_{\text{nc}} (\eta; T) &= S_{\text{nc}} (\eta; T) -
  S_{\text{nc}} (\eta; T = 0) \notag \\
  &= \tilde S_{\text{nc}} (\eta; T) - \tilde S_{\text{nc}} (\eta; T = 0),
\label{deltaSnc}
}
where the tilde indicates that the regularization is ignored when the action is calculated. The second equality is based on the fact that the regularization is only important at frequencies $\omega \gg T$, but the high-frequency contribution to the action does not depend on~$T$, because $F_0 \to 1$ for $\omega \to \infty$. Equation~\eq{SncT} then follows from Eq.~\eq{deltaSnc}

It can be easily seen that the expansion of $\ln\, \langle Z(\eta) \rangle \simeq i S(\eta)$ contains only even powers of~$\eta$. Hence, the odd-order cumulants vanish in thermal equilibrium, which is a consequence of the equivalence of positive and negative direction of the current. The odd-order cumulants are useful for studying the nonequilibrium properties at finite temperatures, because they are not obscured by thermal fluctuations~\cite{gutm03}.

\begin{widetext}

\subsubsection{Transport at zero temperature}

For the medium at zero temperature in the presence of pumping ($F_* > F_0 = 1$) the leading contribution to the action becomes
\eqn{
  i S_0 (\eta) = - \frac {\nu_0 D t_0 \Delta \omega} {4L} (\ln \lambda)^2 =
  - \frac {\nu_0 D t_0 \Delta \omega} {4L} \ln^2 \l\{ 1 - (F_* - 1) 
  (e^{2 i \alpha} - 1) + \sqrt{\l[ 1 - (F_* - 1) 
  (e^{2 i \alpha} - 1) \r]^2 - 1} \r\}.
\label{S0pump}
}
\end{widetext}
Here $\lambda$ is one of the eigenvalues
\aln{
  &\lambda_{1,2} = 1 - Y \pm \sqrt{Y (Y - 2)}, \\
  &Y \equiv (F_* - 1) (e^{2 i \alpha} - 1),
}
of the matrix
\eqn{
  \hat B \equiv \hat \Lambda_* \, e^{-i \alpha \hat \gamma}
  \hat \Lambda_0 \, e^{i \alpha \hat \gamma}.
}
It is convenient to calculate the nonconservative part~\eq{Snc} by diagonalizing the matrix~$\hat J_0$. This is achieved with the transformation $\hat {\mc R}^{-1} \hat J_0 \hat {\mc R}$, where the rotation matrices
\eqn{
  \hat {\mc R} = \big(\hat R^{(1)}, \hat R^{(2)} \big), \quad
  \hat {\mc R}^{-1} = \l(\begin{array}{c} (\hat L^{(1)})^\dag \\
  (\hat L^{(2)})^\dag \end{array} \r)
}
are defined by the biorthogonal right and left eigenvectors of matrix~$\hat B$, $\hat R^{(j)}$ and $\hat L^{(j)}$, corresponding to the eigenvalue~$\lambda_j$. After some algebra we arrive~at
\aln{
  &i S_{\text{nc}} (\eta) = - \frac {\nu_0 t_0 \Delta \omega L}
  {4 \tau_{\text a}} \notag \\
  &\times \sum_{j=1}^2 \l\{ \frac {\lambda_j - 1} {\ln \lambda_j}
  \l[1 + 2 (F_* - 1) \l(L^{(j)}_1\r)^* R^{(j)}_2 \r] - 1 \r\}\!,
\label{Snc_pump}
}
where, again, the regularization makes it necessary to subtract unity.

The linear contribution in the $S (\eta)$ expansion yields the first-order cumulant, the average transmitted energy, which is proportional to the average current
\eqn{
  \langle j \rangle \simeq \langle j_0 \rangle \l( 1 - \frac {L^2}
  {6 l_{\text a}^2} \r).
\label{j_av}
}
The average current in conservative medium,
\eqn{
  \langle j_0 \rangle = \nu_0 \, \Delta \omega\, \omega_0 D n_* L^{-1},
\label{j0}
}
is linear with the photon occupation number (pump intensity) $n_* = (F_* - 1)/2$ and inversely proportional to the length of the channel. To clarify the meaning of these expressions, we derive them, alternatively, from the nonconservative diffusion equation~\eq{difeq}, which, in the stationary regime, takes the form
\eqn{
  \pd_x^2 n - l_{\text a}^{-2} n = 0.
}
The solution for the occupation number satisfying the boundary conditions $n(0) = n_*$ and $n(L) = 0$~is
\eqn{
  n(x) = n_* \frac {\sinh\, [(L - x)/l_{\text a}]} {\sinh\, (L/ l_{\text a})}.
}
Expanding the energy current $j(L) = -\omega_0 D n' (L)$ in $L/l_{\text a}$ we obtain Eqs.~\eq{j_av} and~\eq{j0} (up to the number of modes~$\nu_0 \, \Delta \omega$).

Higher-order cumulants can be calculated by expanding Eqs.~\eq{S0pump} and~\eq{Snc_pump} in $i \eta$, possibly, with the help of a symbolic manipulation software. Specifically, we find
\aln{
  &\frac {C_2/\omega_0^2} {C_1^{(0)} \! /\omega_0} \simeq \frac 2 3 n_*
  \l(1 - \frac {11} {10} \frac {L^2} {l_{\text a}^2} \r),
\label{C2} \\
  &\frac {C_3/\omega_0^3} {C_1^{(0)} \! /\omega_0} \simeq \frac {16} {15}
  n_*^2 \l(1 - \frac {191} {336} \frac {L^2} {l_{\text a}^2} \r),
\label{C3}
}
where $C_1^{(0)} = \langle j_0 \rangle\, t_0$ and we take into account that \mbox{$n_* \gg 1$} for a classical electromagnetic wave. The ratios $C_m/ \omega_0^m$ describe fluctuations of the number of transmitted photons $E/\omega_0$. The principal terms in Eqs.~\eq{C2} and~\eq{C3} agree with the results of diagrammatic calculation for the square pump profile incident on a disordered slab~\cite{debo94}. The absorption (amplification) leads to decrease (increase) of fluctuations.

The results for the conservative medium ($l_{\text a}^{-2} = 0$) are in agreement with the Dorokhov's distribution~\cite{doro84} of the transmission coefficients,
\eqn{
  P (\mc T) \simeq \frac {P_0} {\mc T \sqrt{1 - \mc T}},
}
for sufficiently narrow pumping bandwidth~$\Delta \omega$ and short measuring time~$t_0$. The distribution was used to derive the average shot noise for electron transport~\cite{been92}. $P (\mc T)$~needs to be regularized at $\mc T \to 0$ to be normalizable, but can be used directly to calculate the moments of~$\mc T$ and, hence, of the transmitted energy $E = n_*\, \omega_0 \mc T$. The constant
\eqn{
  P_0 = \frac {\nu_0 \Delta \omega D t_0} {2L}
}
can be determined by comparing the average $\langle E  \rangle = n_*\, \omega_0 \langle \mc T \rangle$ with $C_1^{(0)}$. Then, for $P_0 \ll 1$, averaging with~$P (\mc T)$ reproduces the leading-order terms in Eqs.~\eq{C2} and~\eq{C3}. After estimating the one-dimensional density of modes as $\nu_0 \sim 1$, we can recast the condition of small~$P_0$ in the form
\eqn{
  \Delta k l_0\, \frac {l_0} L \sim \Delta k L \frac {t_0}
  {t_{\text{Th}}} \ll 1,
}
where $\Delta k = \Delta \omega$ is the wave number pumping bandwidth, $l_0 = \sqrt{D t_0}$ is the distance that the energy diffuses in time~$t_0$, and the Thouless time $t_{\text{Th}} = L^2/D$ is the time that it takes to diffuse the distance~$L$.

Despite the similarities, the fluctuations of transmitted energy of the classical wave and the fluctuations of transmitted electron charge (shot noise) are of different origin. In the latter case, the number of transmitted electrons fluctuates for a given realization of~$\mc T$, and the average noise is obtained after integrating with~$P (\mc T)$. In the classical case, the energy fluctuations result from the fluctuations of~$\mc T$, whereas no fluctuations occur for a given disorder realization.

\mbox{}

\section{Conclusions}

In the framework of Keldysh nonlinear $\sigma$ model for classical electromagnetic waves in nonconservative disordered medium we derived an action that includes the source term for the energy current. Within the stationary-phase approximation we obtained a generating function for the cumulants of the energy transmitted through the weakly nonconservative one-dimensional disordered system. The odd-order cumulants for thermal fluctuations vanish in the absence of pumping, which is a consequence of the symmetry of the system. In the pumped system the fluctuations over the ensemble of disorder realizations can be related to the Dorokhov's distribution of transmition coefficients. Our results for a conservative medium quantitatively agree with previous diagrammatic calculations of the low-order cumulants. Moreover, we show that the absorption (amplification) causes reduction (increase) of fluctuations. The photon concentration, or energy density, in the nonconservative medium is shown to obey the diffusion equation with relaxation term; the gradient of the concentration determines the average current. 

\emph{Note added in proof.} Photon shot noise in a nonconservative
medium was considered in Refs.~\cite{been99,patr99}.

\bibliography{transport}

\begin{thebibliography}{10}%
\makeatletter
\providecommand \@ifxundefined [1]{%
 \ifx #1\undefined \expandafter \@firstoftwo
 \else \expandafter \@secondoftwo
\fi
}%
\providecommand \@ifnum [1]{%
 \ifnum #1\expandafter \@firstoftwo
 \else \expandafter \@secondoftwo
\fi
}%
\providecommand \enquote [1]{``#1''}%
\providecommand \bibnamefont  [1]{#1}%
\providecommand \bibfnamefont [1]{#1}%
\providecommand \citenamefont [1]{#1}%
\providecommand\href[0]{\@sanitize\@href}%
\providecommand\@href[1]{\endgroup\@@startlink{#1}\endgroup\@@href}%
\providecommand\@@href[1]{#1\@@endlink}%
\providecommand \@sanitize [0]{\begingroup\catcode`\&12\catcode`\#12\relax}%
\@ifxundefined \pdfoutput {\@firstoftwo}{%
 \@ifnum{\z@=\pdfoutput}{\@firstoftwo}{\@secondoftwo}%
}{%
 \providecommand\@@startlink[1]{\leavevmode\special{html:<a href="#1">}}%
 \providecommand\@@endlink[0]{\special{html:</a>}}%
}{%
 \providecommand\@@startlink[1]{%
  \leavevmode
  \pdfstartlink
   attr{/Border[0 0 1 ]/H/I/C[0 1 1]}%
   user{/Subtype/Link/A<</Type/Action/S/URI/URI(#1)>>}%
  \relax
 }%
 \providecommand\@@endlink[0]{\pdfendlink}%
}%
\providecommand \url  [0]{\begingroup\@sanitize \@url }%
\providecommand \@url [1]{\endgroup\@href {#1}{\urlprefix}}%
\providecommand \urlprefix [0]{URL }%
\providecommand \Eprint[0]{\href }%
\@ifxundefined \urlstyle {%
  \providecommand \doi [1]{doi:\discretionary{}{}{}#1}%
}{%
  \providecommand \doi [0]{doi:\discretionary{}{}{}\begingroup
  \urlstyle{rm}\Url }%
}%
\providecommand \doibase [0]{http://dx.doi.org/}%
\providecommand \Doi[1]{\href{\doibase#1}}%
\providecommand \bibAnnote [3]{%
  \BibitemShut{#1}%
  \begin{quotation}\noindent
    \textsc{Key:}\ #2\\\textsc{Annotation:}\ #3%
  \end{quotation}%
}%
\providecommand \bibAnnoteFile [2]{%
  \IfFileExists{#2}{\bibAnnote {#1} {#2} {\input{#2}}}{}%
}%
\providecommand \typeout [0]{\immediate \write \m@ne }%
\providecommand \selectlanguage [0]{\@gobble}%
\providecommand \bibinfo [0]{\@secondoftwo}%
\providecommand \bibfield [0]{\@secondoftwo}%
\providecommand \translation [1]{[#1]}%
\providecommand \BibitemOpen[0]{}%
\providecommand \bibitemStop [0]{}%
\providecommand \bibitemNoStop [0]{.\EOS\space}%
\providecommand \EOS [0]{\spacefactor3000\relax}%
\providecommand \BibitemShut [1]{\csname bibitem#1\endcsname}%
\bibitem{blan00}%
  \BibitemOpen
  \bibfield{author}{%
  \bibinfo {author} {\bibfnamefont{Y.~M.}\ \bibnamefont{Blanter}}\ and\
  \bibinfo {author} {\bibnamefont{B{\"u}ttiker}},\ }%
  \bibfield{journal}{%
  \bibinfo {journal} {Phys. Rep.}\ }%
  \textbf{\bibinfo {volume} {336}},\ \bibinfo {pages} {1} (\bibinfo {year}
  {2000})%
  \bibAnnoteFile{NoStop}{blan00}%
\bibitem{been03}%
  \BibitemOpen
  \bibfield{author}{%
  \bibinfo {author} {\bibfnamefont{C.~W.~J.}\ \bibnamefont{Beenakker}}\ and\
  \bibinfo {author} {\bibnamefont{Sch{\"o}nenberger}},\ }%
  \bibfield{journal}{%
  \bibinfo {journal} {Phys. Today}\ }%
  \textbf{\bibinfo {volume} {56}},\ \bibinfo {pages} {37} (\bibinfo {year}
  {2003})%
  \bibAnnoteFile{NoStop}{been03}%
\bibitem{naza09}%
  \BibitemOpen
  \bibfield{author}{%
  \bibinfo {author} {\bibfnamefont{Y.~V.}\ \bibnamefont{Nazarov}}\ and\
  \bibinfo {author} {\bibfnamefont{Y.~M.}\ \bibnamefont{Blanter}},\ }%
  \emph{\bibinfo {title} {Quantum Transport: Introduction to Nanoscience}}\
  (\bibinfo {publisher} {Cambridge University Press},\ \bibinfo {address} {New
  York},\ \bibinfo {year} {2009})%
  \bibAnnoteFile{NoStop}{naza09}%
\bibitem{naza03}%
  \BibitemOpen
  \bibinfo {editor} {\bibfnamefont{Y.~V.}\ \bibnamefont{Nazarov}},\ ed.,\
  \emph{\bibinfo {title} {Quantum Noise in Mesoscopic Physics}},\ NATO Science
  Series\ (\bibinfo {publisher} {Kluwer Academic Publishers},\ \bibinfo
  {address} {Dordrecht},\ \bibinfo {year} {2002})%
  \bibAnnoteFile{NoStop}{naza03}%
\bibitem{lern04}%
  \BibitemOpen
  \bibinfo {editor} {\bibfnamefont{I.~V.}\ \bibnamefont{Lerner}}, \bibinfo
  {editor} {\bibfnamefont{B.~L.}\ \bibnamefont{Altshuler}},\ and\ \bibinfo
  {editor} {\bibfnamefont{Y.}~\bibnamefont{Gefen}},\ eds.,\ \emph{\bibinfo
  {title} {Fundamental Problems of Mesoscopic Physics: Interactions and
  Decoherence}},\ NATO Science Series\ (\bibinfo {publisher} {Kluwer Academic
  Publishers},\ \bibinfo {address} {Dordrecht},\ \bibinfo {year} {2004})%
  \bibAnnoteFile{NoStop}{lern04}%
\bibitem{jong96}%
  \BibitemOpen
  \bibfield{author}{%
  \bibinfo {author} {\bibfnamefont{M.~J.~M.}\ \bibnamefont{de~Jong}}\ and\
  \bibinfo {author} {\bibfnamefont{C.~W.~J.}\ \bibnamefont{Beenakker}},\ }%
  \bibfield{journal}{%
  \bibinfo {journal} {Physica A}\ }%
  \textbf{\bibinfo {volume} {230}},\ \bibinfo {pages} {219} (\bibinfo {year}
  {1996})%
  \bibAnnoteFile{NoStop}{jong96}%
\bibitem{been92}%
  \BibitemOpen
  \bibfield{author}{%
  \bibinfo {author} {\bibfnamefont{C.~W.~J.}\ \bibnamefont{Beenakker}}\ and\
  \bibinfo {author} {\bibfnamefont{M.}~\bibnamefont{B{\"u}ttiker}},\ }%
  \bibfield{journal}{%
  \bibinfo {journal} {Phys. Rev. B}\ }%
  \textbf{\bibinfo {volume} {46}},\ \bibinfo {pages} {1889} (\bibinfo {year}
  {1992})%
  \bibAnnoteFile{NoStop}{been92}%
\bibitem{naza94}%
  \BibitemOpen
  \bibfield{author}{%
  \bibinfo {author} {\bibfnamefont{Y.~V.}\ \bibnamefont{Nazarov}},\ }%
  \bibfield{journal}{%
  \bibinfo {journal} {Phys. Rev. Lett.}\ }%
  \textbf{\bibinfo {volume} {73}},\ \bibinfo {pages} {134} (\bibinfo {year}
  {1994})%
  \bibAnnoteFile{NoStop}{naza94}%
\bibitem{doro84}%
  \BibitemOpen
  \bibfield{author}{%
  \bibinfo {author} {\bibfnamefont{O.~N.}\ \bibnamefont{Dorokhov}},\ }%
  \bibfield{journal}{%
  \bibinfo {journal} {Solid State Commun.}\ }%
  \textbf{\bibinfo {volume} {51}},\ \bibinfo {pages} {381} (\bibinfo {year}
  {1984})%
  \bibAnnoteFile{NoStop}{doro84}%
\bibitem{gena87}%
  \BibitemOpen
  \bibfield{author}{%
  \bibinfo {author} {\bibfnamefont{A.~Z.}\ \bibnamefont{Genack}},\ }%
  \bibfield{journal}{%
  \bibinfo {journal} {Phys. Rev. Lett.}\ }%
  \textbf{\bibinfo {volume} {58}},\ \bibinfo {pages} {2043} (\bibinfo {year}
  {1987})%
  \bibAnnoteFile{NoStop}{gena87}%
\bibitem{vell08}%
  \BibitemOpen
  \bibfield{author}{%
  \bibinfo {author} {\bibfnamefont{I.~M.}\ \bibnamefont{Vellekoop}}\ and\
  \bibinfo {author} {\bibfnamefont{A.~P.}\ \bibnamefont{Mosk}},\ }%
  \bibfield{journal}{%
  \bibinfo {journal} {Phys. Rev. Lett.}\ }%
  \textbf{\bibinfo {volume} {101}},\ \bibinfo {pages} {120601} (\bibinfo {year}
  {2008})%
  \bibAnnoteFile{NoStop}{vell08}%
\bibitem{popo10}%
  \BibitemOpen
  \bibfield{author}{%
  \bibinfo {author} {\bibfnamefont{S.~M.}\ \bibnamefont{Popoff}}, \bibinfo
  {author} {\bibfnamefont{G.}~\bibnamefont{Lerosey}}, \bibinfo {author}
  {\bibfnamefont{R.}~\bibnamefont{Carminati}}, \bibinfo {author}
  {\bibfnamefont{M.}~\bibnamefont{Fink}}, \bibinfo {author}
  {\bibfnamefont{A.~C.}\ \bibnamefont{Boccara}},\ and\ \bibinfo {author}
  {\bibfnamefont{S.}~\bibnamefont{Gigan}},\ }%
  \bibfield{journal}{%
  \bibinfo {journal} {Phys. Rev. Lett.}\ }%
  \textbf{\bibinfo {volume} {104}},\ \bibinfo {pages} {100601} (\bibinfo {year}
  {2010})%
  \bibAnnoteFile{NoStop}{popo10}%
\bibitem{shi12}%
  \BibitemOpen
  \bibfield{author}{%
  \bibinfo {author} {\bibfnamefont{Z.}~\bibnamefont{Shi}}\ and\ \bibinfo
  {author} {\bibfnamefont{A.~Z.}\ \bibnamefont{Genack}},\ }%
  \bibfield{journal}{%
  \bibinfo {journal} {Phys. Rev. Lett.}\ }%
  \textbf{\bibinfo {volume} {108}},\ \bibinfo {pages} {043901} (\bibinfo {year}
  {2012})%
  \bibAnnoteFile{NoStop}{shi12}%
\bibitem{davy13}%
  \BibitemOpen
  \bibfield{author}{%
  \bibinfo {author} {\bibfnamefont{M.}~\bibnamefont{Davy}}, \bibinfo {author}
  {\bibfnamefont{Z.}~\bibnamefont{Shi}}, \bibinfo {author}
  {\bibfnamefont{J.}~\bibnamefont{Wang}},\ and\ \bibinfo {author}
  {\bibfnamefont{A.~Z.}\ \bibnamefont{Genack}},\ }%
  \bibfield{journal}{%
  \bibinfo {journal} {Opt. Express}\ }%
  \textbf{\bibinfo {volume} {21}},\ \bibinfo {pages} {10367} (\bibinfo {year}
  {2013})%
  \bibAnnoteFile{NoStop}{davy13}%
\bibitem{debo94}%
  \BibitemOpen
  \bibfield{author}{%
  \bibinfo {author} {\bibfnamefont{J.~F.}\ \bibnamefont{de~Boer}}, \bibinfo
  {author} {\bibfnamefont{M.~C.~W.}\ \bibnamefont{van Rossum}}, \bibinfo
  {author} {\bibfnamefont{M.~P.}\ \bibnamefont{van Albada}}, \bibinfo {author}
  {\bibfnamefont{T.~M.}\ \bibnamefont{Nieuwenhuizen}},\ and\ \bibinfo {author}
  {\bibfnamefont{A.}~\bibnamefont{Lagendijk}},\ }%
  \bibfield{journal}{%
  \bibinfo {journal} {Phys. Rev. Lett.}\ }%
  \textbf{\bibinfo {volume} {73}},\ \bibinfo {pages} {2567} (\bibinfo {year}
  {1994})%
  \bibAnnoteFile{NoStop}{debo94}%
\bibitem{vanl96}%
  \BibitemOpen
  \bibfield{author}{%
  \bibinfo {author} {\bibfnamefont{S.~A.}\ \bibnamefont{van Langen}}, \bibinfo
  {author} {\bibfnamefont{P.~W.}\ \bibnamefont{Brouwer}},\ and\ \bibinfo
  {author} {\bibfnamefont{C.~W.~J.}\ \bibnamefont{Beenakker}},\ }%
  \bibfield{journal}{%
  \bibinfo {journal} {Phys. Rev. E}\ }%
  \textbf{\bibinfo {volume} {53}},\ \bibinfo {pages} {R1344} (\bibinfo {year}
  {1996})%
  \bibAnnoteFile{NoStop}{vanl96}%
\bibitem{cao05}%
  \BibitemOpen
  \bibfield{author}{%
  \bibinfo {author} {\bibfnamefont{H.}~\bibnamefont{Cao}},\ }%
  \bibfield{journal}{%
  \bibinfo {journal} {J. Phys. A}\ }%
  \textbf{\bibinfo {volume} {38}},\ \bibinfo {pages} {10497} (\bibinfo {year}
  {2005})%
  \bibAnnoteFile{NoStop}{cao05}%
\bibitem{lai12}%
  \BibitemOpen
  \bibfield{author}{%
  \bibinfo {author} {\bibfnamefont{Z.~Y.}\ \bibnamefont{Lai}}\ and\ \bibinfo
  {author} {\bibfnamefont{O.}~\bibnamefont{Zaitsev}},\ }%
  \bibfield{journal}{%
  \bibinfo {journal} {Phys. Rev. A}\ }%
  \textbf{\bibinfo {volume} {85}},\ \bibinfo {pages} {043838} (\bibinfo {year}
  {2012})%
  \bibAnnoteFile{NoStop}{lai12}%
\bibitem{kame11}%
  \BibitemOpen
  \bibfield{author}{%
  \bibinfo {author} {\bibfnamefont{A.}~\bibnamefont{Kamenev}},\ }%
  \emph{\bibinfo {title} {Field Theory of Non-Equilibrium Systems}}\ (\bibinfo
  {publisher} {Cambridge University Press},\ \bibinfo {address} {Cambridge},\
  \bibinfo {year} {2011})%
  \bibAnnoteFile{NoStop}{kame11}%
\bibitem{keld64a}%
  \BibitemOpen
  \bibfield{author}{%
  \bibinfo {author} {\bibfnamefont{L.~V.}\ \bibnamefont{Keldysh}},\ }%
  \bibfield{journal}{%
  \bibinfo {journal} {Zh.\ Eksp.\ Teor.\ Fiz.}\ }%
  \textbf{\bibinfo {volume} {47}},\ \bibinfo {pages} {1515} (\bibinfo {year}
  {1964}),\ \bibinfo {note} {in Russian}%
  \bibAnnoteFile{NoStop}{keld64a}%
\bibitem{keld64b}%
  \BibitemOpen
  \bibfield{journal}{%
  \bibinfo {journal} {Sov.\ Phys.\ JETP}\ }%
  \textbf{\bibinfo {volume} {20}},\ \bibinfo {pages} {1018} (\bibinfo {year}
  {1965})%
  \bibAnnoteFile{NoStop}{keld64b}%
\bibitem{Note1}%
  \BibitemOpen
  \bibinfo {note} {In Ref.~\cite {lai12} we did not discriminate between
  $\protect \mathaccentV {hat}05E\Lambda $ and~$\protect \mathaccentV
  {hat}05E\Lambda _0$, hence, the results of this reference are valid only in
  the case of thermal equilibrium.}%
  \bibAnnoteFile{Stop}{Note1}%
\bibitem{Note2}%
  \BibitemOpen
  \bibinfo {note} {Note that $j^\protect \text {cl}/ \protect \sqrt 2 = (j_+ +
  j_-)/2$}%
  \bibAnnoteFile{NoStop}{Note2}%
\bibitem{Note3}%
  \BibitemOpen
  \bibinfo {note} {To regularize the integral at low frequencies we should
  avoid fixing~$\omega $ in $\alpha = \omega \eta /2$.}%
  \bibAnnoteFile{Stop}{Note3}%
\bibitem{gutm03}%
  \BibitemOpen
  \bibfield{author}{%
  \bibinfo {author} {\bibfnamefont{D.~B.}\ \bibnamefont{Gutman}}, \bibinfo
  {author} {\bibfnamefont{Y.}~\bibnamefont{Gefen}},\ and\ \bibinfo {author}
  {\bibfnamefont{A.~D.}\ \bibnamefont{Mirlin}},\ }%
  in\ \emph{\bibinfo {booktitle} {Quantum Noise in Mesoscopic Physics}},\
  \bibinfo {series} {NATO Science Series}, Vol.~\bibinfo {volume} {97},\
  \bibinfo {editor} {edited by\ \bibinfo {editor} {\bibfnamefont{Y.~V.}\
  \bibnamefont{Nazarov}}}\ (\bibinfo {publisher} {Springer Netherlands},\
  \bibinfo {year} {2003})\ p.\ \bibinfo {pages} {497}%
  \bibAnnoteFile{NoStop}{gutm03}%
\bibitem{been99}%
  \BibitemOpen
  \bibfield{author}{%
  \bibinfo {author} {\bibfnamefont{C.~W.~J.}\ \bibnamefont{Beenakker}}\ and\
  \bibinfo {author} {\bibfnamefont{M.}~\bibnamefont{Patra}},\ }%
  \bibfield{journal}{%
  \bibinfo {journal} {Mod. Phys. Lett. B}\ }%
  \textbf{\bibinfo {volume} {13}},\ \bibinfo {pages} {337} (\bibinfo {year}
  {1999})%
  \bibAnnoteFile{NoStop}{been99}%
\bibitem{patr99}%
  \BibitemOpen
  \bibfield{author}{%
  \bibinfo {author} {\bibfnamefont{M.}~\bibnamefont{Patra}}\ and\ \bibinfo
  {author} {\bibfnamefont{C.~W.~J.}\ \bibnamefont{Beenakker}},\ }%
  \bibfield{journal}{%
  \bibinfo {journal} {Phys. Rev. A}\ }%
  \textbf{\bibinfo {volume} {60}},\ \bibinfo {pages} {4059} (\bibinfo {year}
  {1999})%
  \bibAnnoteFile{NoStop}{patr99}%
\end{thebibliography}%

\end{document}